\documentclass[11pt,twoside]{article}

\usepackage{surace_asp2004}
\usepackage{epsf}
\usepackage{epsfig}
\usepackage{lscape}

\begin{document}
\title{{\it Spitzer} Mid-Infrared Imaging of Nearby Ultraluminous Infrared Galaxies}   
\author{J.A. Surace, Z. Wang, S. Willner, H. Smith, J. Pipher, W. Forrest, 
and G. Fazio}  
\affil{Spitzer Science Center/Caltech, Harvard-Smithsonian Center for 
Astrophysics, 
University of Rochester} 

\begin{abstract} 
    We have observed 14 nearby ($z<0.16$) Ultraluminous Infrared
    Galaxies (ULIRGs) with {\it Spitzer} at 3.6---24\micron.  The
    underlying host galaxies are well-detected, in addition to the
    luminous nuclear cores. While the spatial resolution of {\it 
    Spitzer} is poor, the great sensitivity of the data reveals the 
    underlying galaxy merger remnant, and provides the first look at 
    off-nuclear mid-infrared activity.
\end{abstract}



\noindent{\bf 1. Background}
\vspace{0.08 truein}

\noindent ULIRGs as a class are characterized by L$_{ir} >
10^{12}${\hbox{{\it L}$_\odot$\ }}, equivalent to the luminosity of classical optically
selected quasars.  Based on extensive ground and space-based imaging,
nearly all are known to be advanced merger remnants resulting from
collisions of gas-rich galaxies.  Dynamics of the collision result in
a rearrangement of the gas, dust, and stellar content of the galaxies.
The rapid inflow of gas deep into the merger core results in a
powerful burst of star formation, and/or the fueling (in at least
one-quarter) of an active nucleus.  It has been postulated that ULIRGs
are the progenitors of both QSOs and elliptical galaxies.  Extensive
imaging campaigns in the UV, optical, near and mid-IR during the '90s
revealed the complex composite nature of the ULIRGs (Surace et al.
1998, 1999,
2000, 2000b, Scoville et al.  2000, Soifer et al.  2000, Goldader
et al.  2002).  At optical/UV wavelengths the emission is dominated by
the old stellar population, as well as actively star-forming ``super
star clusters'' located both in the nuclear regions and along the
tidal features.  At infrared wavelengths the emission is increasingly
due to thermal dust emission arising in a compact core typically less
than 100pc in diameter.  Ground-based mid-infrared observations
constrained a large fraction of the IRAS 12 and 25\micron \
emission to an unresolved core.

\vspace{0.12 truein}

\noindent{\bf 2. {\it Spitzer} Mid-IR Imaging}
\vspace{0.08 truein}

\noindent As part of the IRAC GTO program, we observed 14 nearby
ULIRGs selected from the well-studied Bright Galaxy and Warm Galaxy
Samples of Sanders et al.  (1988, 1988b).  The spatial resolution of Spitzer
is relatively poor ($\approx$2\arcsec) at IRAC wavelengths compared to
the known optical compact structure in these galaxies.  However, it
can resolve the galaxy bodies and extended tidal features, and
separate the known double nuclei in some systems.  While 
ground-based observations (e.g. from Keck) offer several times higher
spatial resolution, they lack the sensitivity to 
detect anything beyond the high surface brightness nuclei.
The extended, and hence potentially resolvable,
emission is too faint to be readily detected from the ground.  We
therefore specifically selected systems that had optical emission on
spatial scales of several arcseconds or more.

\begin{figure}
    \centering \leavevmode
        \epsfxsize=.8\textwidth \epsfbox{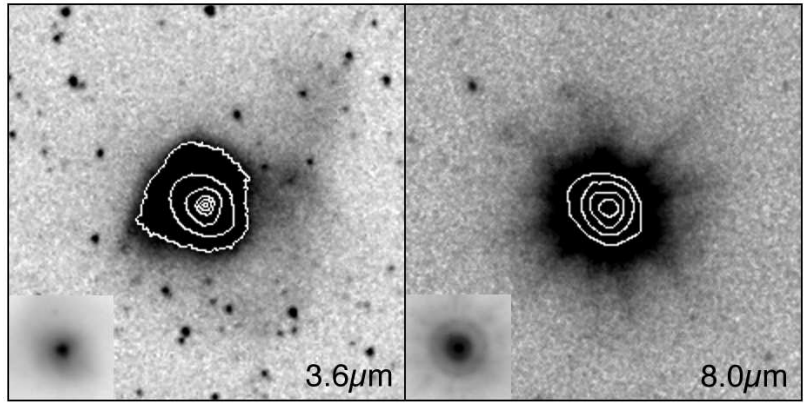}
    \caption{{\bf Arp 220}, the nearest ULIRG. Shown with a logarithmic 
    scale, the 3.6\micron \ image clearly shows the extended galaxy body. 
    At 8\micron \ the underlying emission from the host galaxy is 
    detected as a faint extended component, but the system is
    increasingly dominated by a high surface brightness region 
    unresolved by {\it Spitzer} that contributes 
    $\approx$2/3 of the total flux in the IRAC filter.
    The small insets show
    a linear stretch of the point-like nuclear core, at the same
    spatial scale as the larger images.}
    \end{figure}
    
\begin{figure}
    \centering \leavevmode
        \epsfxsize=.8\textwidth \epsfbox{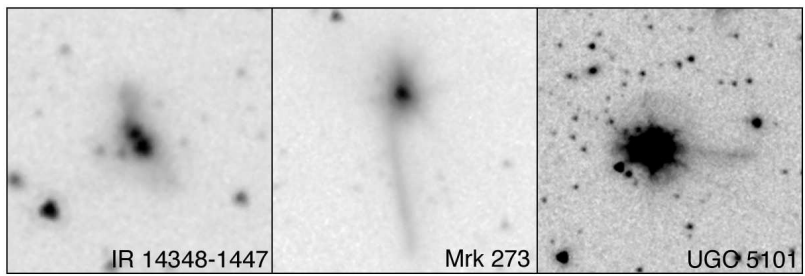}
    \caption{3.6\micron \ images of three selected ``cool'' ULIRGs. The galaxy 
    bodies and tidal features, composed of stars, are clearly visible at 
    this wavelength as are one or more compact nuclei. In the more
    AGN-like ``warm'' ULIRGs (Sanders et al. 1988b) the galaxies are
    more dominated by the nuclear core, a result seen previously at
    K-band.}
    \end{figure}
    
All of the galaxy bodies are well-detected at the shorter 
wavelengths, and appear very similar to previous optical and near-IR 
imaging. There is little evidence for extended emission (e.g. tidal 
features) beyond that 
previously known. At longer wavelengths the ULIRGs are increasingly dominated by the 
nuclear core, which is generally unresolved by {\it Spitzer}.  
Preliminary analysis suggests that the cores of the double nucleus 
systems appear more diffuse than the single nucleus systems. Further 
analysis will examine the colors of the galaxy bodies, in order to 
characterize the dust distribution and the state of star formation in the 
merger remnants.

\acknowledgements
JAS was supported by the Jet Propulsion Laboratory, California
Institute of Technology, under contract with NASA.




\end{document}